\documentclass[prx,showpacs,twocolumn, preprintnumbers,
amsmath,amssymb,superscriptaddress,nofootinbib,longbibliography,nofootinbibjustification=justified,singlelinecheck=false]{revtex4-2}

\usepackage{amsfonts}
\usepackage{graphicx}
\usepackage{hyperref} 
\usepackage{cleveref}
\hypersetup{
    hidelinks,
    colorlinks=true,
    breaklinks=true,
    citecolor=SteelBlue,
    filecolor=LimeGreen,
    linkcolor=MediumBlue,
    urlcolor=MediumPurple
}
 
\usepackage[svgnames, table]{xcolor}
\usepackage{xspace}
\usepackage{multirow}
\usepackage{array}
\usepackage{mathtools}
\usepackage[normalem]{ulem}
\usepackage{booktabs}

\renewcommand{\arraystretch}{1.}  
\graphicspath{{./figures/}}

\usepackage{tikz}
\usetikzlibrary{calc}

\definecolor{blueA}{RGB}{178,193,218}
\definecolor{blueB}{RGB}{149,181,214}
\definecolor{blueC}{RGB}{76,160,224}
\definecolor{greenQ}{RGB}{173,206,92}
\definecolor{greenL}{RGB}{191,199,194}
\definecolor{tealCap}{RGB}{138,191,184}

\tikzset{
  nodeblack/.style={circle,fill=black,inner sep=1.9pt},
  nodeblue/.style={circle,draw=blue!70!black,fill=blue!25,inner sep=1.7pt,line width=0.8pt},
  edgeblack/.style={line width=0.9pt,black},
  edgeblue/.style={line width=0.9pt,blue!70!black}
}

\newcommand{\gIso}{%
\begin{tikzpicture}[baseline=-0.5ex,scale=0.55]
  \node[nodeblack] at (0,0) {};
\end{tikzpicture}
}

\newcommand{\gEdge}{%
\begin{tikzpicture}[baseline=-0.5ex,scale=0.55]
  \draw[edgeblack] (0,0)--(0.8,0);
  \node[nodeblack] at (0,0) {};
  \node[nodeblack] at (0.8,0) {};
\end{tikzpicture}
}

\newcommand{\gThreeBlue}{%
\begin{tikzpicture}[baseline=-0.5ex,scale=0.55]
  \coordinate (a) at (0.4,0.75);
  \coordinate (b) at (0,0);
  \coordinate (c) at (0.8,0);
  \draw[edgeblue] (a)--(b) (a)--(c);
  \node[nodeblue] at (a) {};
  \node[nodeblue] at (b) {};
  \node[nodeblue] at (c) {};
\end{tikzpicture}
}

\newcommand{\gStarFour}{%
\begin{tikzpicture}[baseline=-0.5ex,scale=0.55]
  \coordinate (a) at (0.4,0.8);
  \coordinate (b) at (0,0);
  \coordinate (c) at (0.4,0);
  \coordinate (d) at (0.8,0);
  \draw[edgeblack] (a)--(b) (a)--(c) (a)--(d);
  \node[nodeblack] at (a) {};
  \node[nodeblack] at (b) {};
  \node[nodeblack] at (c) {};
  \node[nodeblack] at (d) {};
\end{tikzpicture}
}

\newcommand{\gTriSquare}{%
\begin{tikzpicture}[baseline=-0.5ex,scale=0.55]
  \coordinate (a1) at (0,0);
  \coordinate (b1) at (0.55,0);
  \coordinate (c1) at (0.275,0.48);
  \draw[edgeblack] (a1)--(b1)--(c1)--cycle;
  \node[nodeblack] at (a1) {};
  \node[nodeblack] at (b1) {};
  \node[nodeblack] at (c1) {};

  \coordinate (a2) at (0.95,0);
  \coordinate (b2) at (1.50,0);
  \coordinate (c2) at (1.50,0.55);
  \coordinate (d2) at (0.95,0.55);
  \draw[edgeblack] (a2)--(b2)--(c2)--(d2)--cycle;
  \node[nodeblack] at (a2) {};
  \node[nodeblack] at (b2) {};
  \node[nodeblack] at (c2) {};
  \node[nodeblack] at (d2) {};
\end{tikzpicture}
}

\newcommand{\gKFour}{%
\begin{tikzpicture}[baseline=-0.5ex,scale=0.55]
  \coordinate (a) at (0,0);
  \coordinate (b) at (0.8,0);
  \coordinate (c) at (0.8,0.8);
  \coordinate (d) at (0,0.8);
  \draw[edgeblack] (a)--(b)--(c)--(d)--cycle;
  \draw[edgeblack] (a)--(c) (b)--(d);
  \node[nodeblack] at (a) {};
  \node[nodeblack] at (b) {};
  \node[nodeblack] at (c) {};
  \node[nodeblack] at (d) {};
\end{tikzpicture}
}

\makeatletter
\renewcommand{\fnum@figure}{Figure \thefigure}
\makeatother
\makeatletter
\renewcommand{\fnum@table}{Table \thetable}
\makeatother

\newcommand{\ket}[1]{|#1\rangle}

\newcommand{\bra}[1]{ \langle #1 \,|}
\begin{document}


\title{Photonic realization of a subgraph extraction in a quantum random network}

\author{Lijun Xia}
\altaffiliation{These authors contributed equally to this work.}
\affiliation{National Laboratory of Solid-state Microstructures, School of Physics, College of Engineering and Applied Sciences, Collaborative Innovation Center of Advanced Microstructures, Nanjing University, Nanjing 210093, China}
\author{Xuemei Gu}
\altaffiliation{These authors contributed equally to this work.}
\affiliation{Institut für Festkörpertheorie und Optik, Friedrich-Schiller-Universität Jena, Jena, Germany}
\author{Kai Wang}
\email{kai.wang@nju.edu.cn}
\affiliation{National Laboratory of Solid-state Microstructures, School of Physics, College of Engineering and Applied Sciences, Collaborative Innovation Center of Advanced Microstructures, Nanjing University, Nanjing 210093, China}
\author{Leizhen Chen}
\affiliation{National Laboratory of Solid-state Microstructures, School of Physics, College of Engineering and Applied Sciences, Collaborative Innovation Center of Advanced Microstructures, Nanjing University, Nanjing 210093, China}
\author{Mario Krenn}
\email{mario.krenn@uni-tuebingen.de}
\affiliation{Machine Learning in Science Cluster, Department for Computer Science, Faculty of Science, University of Tuebingen, Tuebingen, Germany}
\author{Yan-Qing Lu}
\affiliation{National Laboratory of Solid-state Microstructures, School of Physics, College of Engineering and Applied Sciences, Collaborative Innovation Center of Advanced Microstructures, Nanjing University, Nanjing 210093, China}
\author{Shining Zhu}
\affiliation{National Laboratory of Solid-state Microstructures, School of Physics, College of Engineering and Applied Sciences, Collaborative Innovation Center of Advanced Microstructures, Nanjing University, Nanjing 210093, China}
\author{Xiao-Song Ma}
\email{xiaosong.ma@nju.edu.cn}
\affiliation{National Laboratory of Solid-state Microstructures, School of Physics, College of Engineering and Applied Sciences, Collaborative Innovation Center of Advanced Microstructures, Nanjing University, Nanjing 210093, China}
\affiliation{Synergetic Innovation Center of Quantum Information and Quantum Physics, University of Science and Technology of China, Hefei, Anhui 230026, China}
\affiliation{Hefei National Laboratory, Hefei 230088, China}

\begin{abstract}
Understanding how complex connectivity emerges in networks is a fundamental challenge in classical and quantum science. In classical random networks, complex subgraphs typically require relatively high connection probabilities, whereas quantum random network theory predicts that such structures can arise at a single, lower threshold through entanglement and local operations. Here, using an integrated silicon photonic chip, we experimentally realize a quantum subgraph predicted by quantum random network theory in a four-node quantum random network. Our integrated platform exploits probabilistic photon-pair sources and coherent control of path modes to prepare a structured quantum subgraph through local transformations and postselection, operating in a threshold regime that differs from classical random networks. We verify that the subgraph state exhibits genuine high-dimensional multipartite entanglement across the nodes, providing experimental evidence that quantum entanglement enables connectivity structures beyond classical accessibility.

\end{abstract}

\maketitle


\textit{Introduction --} Networks provide a compact description of connectivity in natural and engineered systems, including the World Wide Web \cite{Adamic2000}, power grids \cite{Watts1998}, neural circuits \cite{wt1999}, and scientific collaborations \cite{doi:10.1126/science.aao0185}. A central problem in network science is to determine how connected subgraphs emerge as the probability of forming links is varied.

The Erdős--Rényi random graph provides a standard setting for studying this question \cite{Erdos1959,Erdos1960}. In this model, a graph of $N$ nodes is formed by connecting each pair of nodes independently with probability $p$. As $p$ increases, different connected structures appear at characteristic probabilities \cite{bollobas1985}. For a specific subgraph, the appearance threshold can be written, up to constants, as $p_c(N,z)\sim N^z$, where the exponent $z$ depends on the subgraph structure. Thus different classical subgraphs generally emerge at different $z$ \cite{Albert2002} (Table~\ref{tab:thresholds}).


Quantum networks extend this setting by allowing links to carry entanglement. Such networks underlie protocols for secure communication \cite{Gisin2007,Kimble2008}, distributed quantum computation \cite{WaltherSci2012,PhysRevLett.132.150604,YCWei2025}, and quantum-enhanced metrology \cite{Komar2014a}. Entanglement also modifies connectivity itself. For example, Acín \textit{et al}. showed that local measurements can establish long-range quantum channels in lattice networks when the initial entanglement exceeds a critical value \cite{Acin2007}. These results motivated the question of how random graph theory changes when graph edges represent probabilistic entangled links rather than classical connections \cite{Perseguers2010a,Meng2021}.


\begin{table}[!t]
\centering
\setlength{\tabcolsep}{8pt}
\renewcommand{\arraystretch}{1.5}
\begin{tabular}{|>{\bfseries}c|c|c|c|c|c|c|}\hline
$z$& \(-\infty\) & \(-2\) & \(-\frac{3}{2}\) & \(-\frac{4}{3}\) & \(-1\)& \(-\frac{2}{3}\) \\
$G_{C}$ & \gIso & \gEdge & \gThreeBlue & \gStarFour & \gTriSquare & \gKFour \\\hline
$z$& \(-\infty\) & \multicolumn{5}{c|}{\(-2\)} \\
$G_{Q}$ & \multicolumn{1}{c|}{\gIso}
        & \multicolumn{1}{c}{\gEdge}
        & \multicolumn{1}{c}{\gThreeBlue}
        & \multicolumn{1}{c}{\gStarFour}
        & \multicolumn{1}{c}{\gTriSquare}
        & \multicolumn{1}{c|}{\gKFour} \\\hline
\end{tabular}
\caption{\textbf{Thresholds for the appearance of classical and quantum subgraphs.} Classical subgraphs \((G_{C})\) appear in random graphs of \(N\) nodes appear at pairwise connection probabilities scaling as \(p\sim N^z\). In quantum random networks, local operations allow finite quantum subgraphs \((G_{Q})\), whose edges represent possible pairs of maximally entangled qubits, to be generated at the common scaling \(z=-2\). The target \(\lambda\) graph highlighted in blue is realized in this work.}
\label{tab:thresholds}
\end{table}
\begin{figure*}[!t]
\centering
\includegraphics[width=0.9\textwidth]{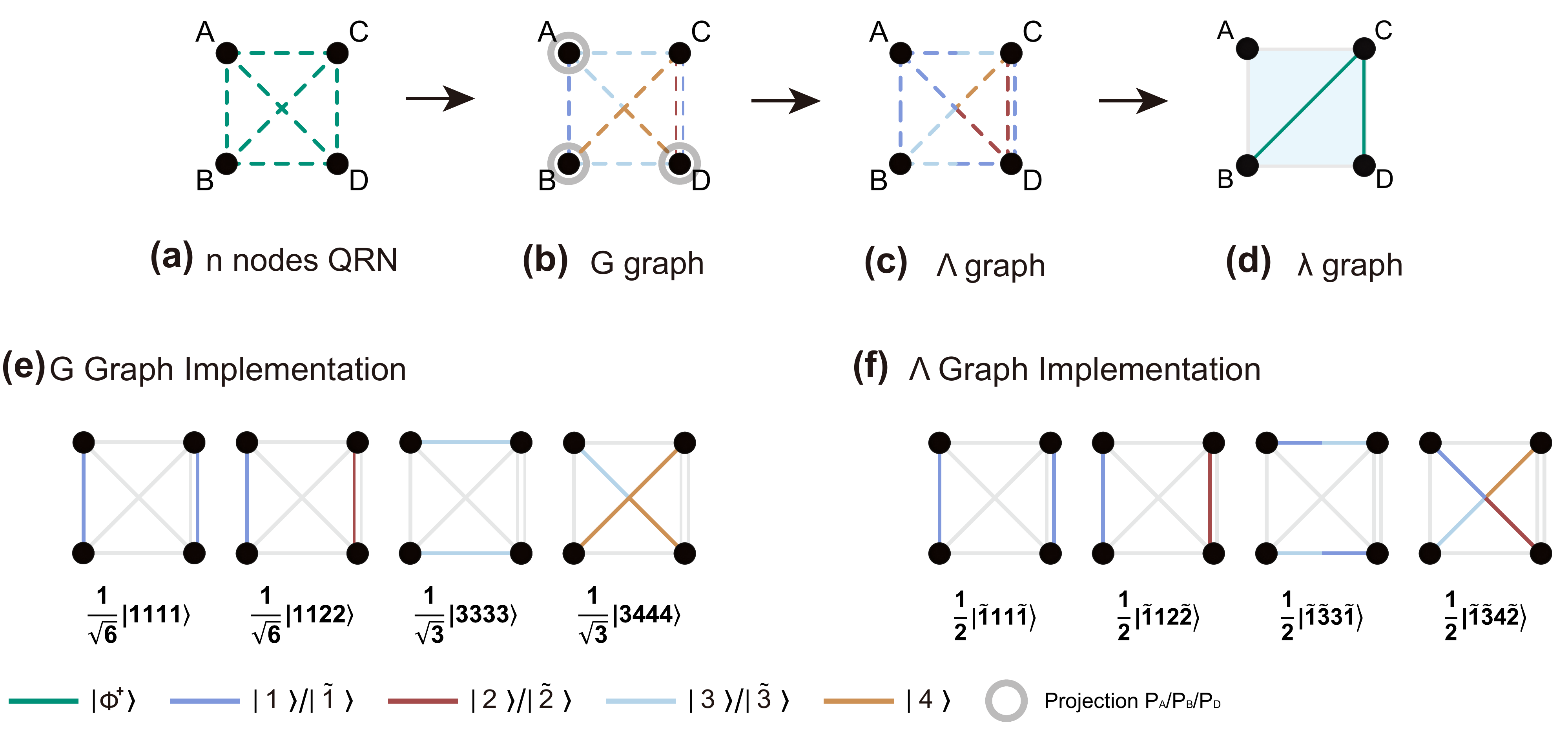}
\begin{center}
\caption{
\textbf{Schematic illustration of the protocol.}
\textbf{(a)} Four-node quantum random network module. Each dashed green edge denotes a probabilistic Bell-state link $\ket{\Phi^+}$.
\textbf{(b)} Experimental pair-creation graph for preparing the resource state $\ket{G}$. Each edge denotes a probabilistic photon-pair creation process, and edge colors label the local photon modes. The two edges between nodes $C$ and $D$ have a relative weight $1/\sqrt{2}$, whereas all other edges have unit weight. The grey circles denote the local transformations $P_A$, $P_B$, and $P_D$.
\textbf{(c)} Graph representation of the postselected state $\ket{\Lambda}$ obtained after the local transformations and postselection on the desired output modes.
\textbf{(d)} Target $\lambda$ graph obtained by recoding node $C$ into two qubits, $C_1$ and $C_2$, yielding $\ket{\lambda}=\ket{\Phi^+}_{BC_1}\otimes\ket{\Phi^+}_{C_2D}$. The solid edges indicate the Bell links represented in the target state after successful postselection and local recoding.
\textbf{(e,f)} Perfect matchings corresponding to the individual terms in $\ket{G}$ and $\ket{\Lambda}$, respectively.}\label{Fig2}
\end{center}
\end{figure*}
This question has been addressed through a quantum extension of random graph theory \cite{Perseguers2010}, which predicts that arbitrary subgraphs can be generated through local operations and classical communication when the connection probability scales as $p \propto N^{-2}$. This scaling contrasts with the classical case and highlights how quantum correlations can lead to distinct connectivity patterns and subgraph thresholds (Table~\ref{tab:thresholds}). Despite these theoretical predictions, experimental verification remains challenging due to substantial resource requirements, particularly on photonic platforms. Proposed photonic implementations encode graph edges using probabilistic photon-pair sources \cite{Krenn2017,Gu2019a,Gu2019,yu2025topo}. Although this graph representation has been experimentally applied to other tasks \cite{Lu2020,Wang2020,Bao2023}, the generation of quantum subgraphs predicted by quantum random network theory has not yet been demonstrated.

In this work, we experimentally demonstrate a finite quantum subgraph in a quantum random network using an integrated silicon photonic chip. Specifically, we realize the $\lambda$ graph highlighted in blue in Table.~\ref{tab:thresholds}, which is a key component of the theoretical prediction in quantum random networks \cite{Perseguers2010}. Our experiment does not directly measure the large-$N$ threshold scaling with network size or connection probability; rather, it realizes the finite local conversion step appearing in the theoretical construction \cite{Perseguers2010}. Our work establishes an integrated photonic
building block for studying quantum network and indicates an important step toward investigating critical behavior in complex quantum networks.

 \begin{figure*}[!t]
\centering
\includegraphics[width=1\textwidth]{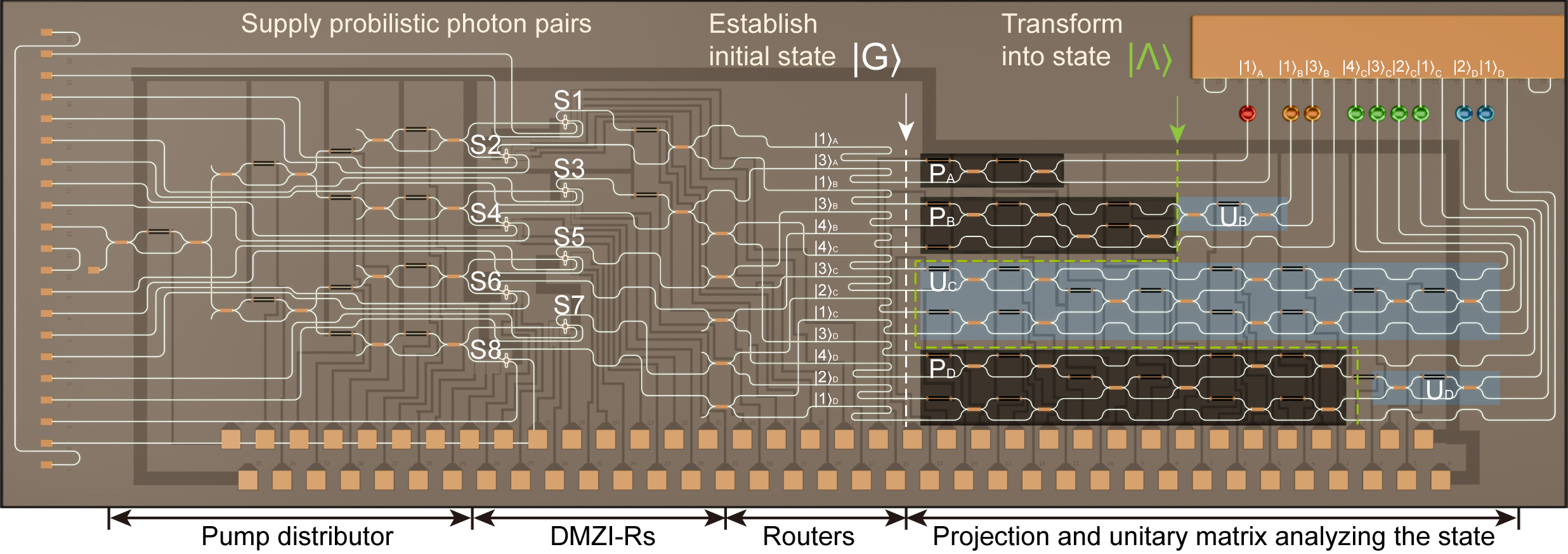}
\begin{center}
\caption{\textbf{Overview of the experiment and the integrated photonic circuit.} Two picosecond pump pulses at different wavelengths are coupled into the chip via a grating coupler and equally distributed to eight DMZI-Rs. Within the DMZI-Rs, the pump pulses (1544.78~nm and 1556.84~nm) generate photon pairs at 1550.81~nm through spontaneous four-wave mixing. The photons are then routed through delay lines that match their arrival times and establish coherent node-to-node connections, preparing the resource state $\ket{G}$ (after the white dashed line). The projections $P_i$ ($i\in\{A,B,D\}$) convert the $\ket{G}$ into $\ket{\Lambda}$ after the green dashed line. Local unitaries $U_i$ ($i\in\{B,C,D\}$) are then applied for quantum-state analysis. Coincidence measurements are performed on the output modes marked by the red, orange, green, and blue spheres (see Supplementary Materials Section~\ref{Scheme} for details). Photons are detected using superconducting nanowire single-photon detectors (SNSPDs), and the electrical signals are processed by an FPGA-based time tagger.}
\label{Fig3}
\end{center}
\end{figure*}

\textit{Quantum Random Network State --}
In a quantum random network, each pair of nodes is supplied, with probability $p$, with a maximally entangled Bell pair $\ket{\Phi^+}$. With joint measurements and local operations, arbitrary finite subgraphs can be generated at the common threshold $p\propto N^{-2}$ in the large-$N$ setting \cite{Perseguers2010}. This differs from classical random graphs, where different subgraphs generally have different appearance thresholds (Table~\ref{tab:thresholds}).

Here we do not realize the full large-$N$ quantum random network. Instead, we implement a finite four-node module of the construction (whose typical subgraphs are listed in Table.\ref{tab:thresholds}), shown in Fig.~\ref{Fig2}(a). The experimental task is to prepare the resource state $\ket{G}$ associated with this module, transform it by fixed local mode operations and postselection into $\ket{\Lambda}$, and verify that $\ket{\Lambda}$ is locally recodable as the target $\lambda$ graph (highlighted in blue in Table.\ref{tab:thresholds}).


For the experimental implementation, it is useful to rewrite the Bell-state edges in Fig.~\ref{Fig2}(a) as photon-pair creation edges in the graph representation introduced in Ref.~\cite{Gu2019}. In this representation, each Bell-state edge in Fig.~\ref{Fig2}(a) is decomposed into two edges, corresponding to the two components of the Bell state. From this expanded set of pair-creation edges, only the edges that contribute to the target four-photon events are retained,
as shown in Fig.~\ref{Fig2}(b). Each edge in Fig.~\ref{Fig2}(b) denotes a photon-pair creation process that creates one photon in each of the two connected nodes. The edge colors label the local photon modes at the two nodes. The numbering of these modes is arbitrary, and indistinguishable modes are assigned the same color \cite{PhysRevLett.118.080401}. The dashed edges represent probabilistic photon-pair sources.

As mentioned above, not all pair-creation edges associated with the Bell-state edges in Fig.~\ref{Fig2}(a) are present in the experimental graph in Fig.~\ref{Fig2}(b): for some Bell-state edges, only one of the two corresponding pair-creation edges is retained. In addition, some retained edges carry reduced weights. In particular, the two edges between nodes $C$ and $D$ have a relative weight $1/\sqrt{2}$, whereas all other retained edges have unit weight. With these mode labels and weights, the graph in Fig.~\ref{Fig2}(b) defines the four-photon resource state $\ket{G}$, which is prepared directly in our experiment.

Conditioning on one photon at each node projects the pair-creation graph in Fig.~\ref{Fig2}(b) onto its perfect matchings. These matchings cover all four nodes exactly once and correspond to the four terms shown in Fig.~\ref{Fig2}(e). After including the relative weights and normalizing, the resource state is
\begin{equation}\label{eq:G}
\ket{G}_{ABCD}
=
\frac{1}{\sqrt{6}}\ket{1111}
+
\frac{1}{\sqrt{6}}\ket{1122}
+
\frac{1}{\sqrt{3}}\ket{3333}
+
\frac{1}{\sqrt{3}}\ket{3444}.
\end{equation}

Starting from $\ket{G}$, we then apply local transformations $P_A$, $P_B$, and $P_D$ to nodes $A$, $B$, and $D$, respectively, as indicated by the grey circles in Fig.~\ref{Fig2}(b). These operations define new local mode bases before a mode-selective postselection. For example, $P_A$ maps $\ket{1}_A$ to $\ket{\tilde{1}}_A=(\ket{1}_A+\ket{3}_A)/\sqrt{2}$. The full definitions are given in Eqs.~\eqref{proj1}--\eqref{proj3}. We then keep only the events with modes $\tilde{1}$ at $A$, modes $1$ or $\tilde{3}$ at $B$, all modes at $C$, and modes $\tilde{1}$ or $\tilde{2}$ at $D$. Conditioned on the prepared resource state $\ket{G}$, this mode-selective postselection gives the state $\ket{\Lambda}$, whose graph representation is shown in Fig.~\ref{Fig2}(c), with success probability $1/6$:
\begin{equation}\label{eq:Lambda_main}
\begin{aligned}
\ket{\Lambda}_{ABCD} &=\frac{1}{2}(\ket{\tilde{1}11\tilde{1}}+\ket{\tilde{1}12\tilde{2}}+\ket{\tilde{1}\tilde{3}3\tilde{1}}+\ket{\tilde{1}\tilde{3}4\tilde{2}})_{ABCD}\\
&=\frac{1}{2}\ket{\tilde{1}}_A(\ket{11\tilde{1}}+\ket{12\tilde{2}}+\ket{\tilde{3}3\tilde{1}}+\ket{\tilde{3}4\tilde{2}})_{BCD}.
\end{aligned}
\end{equation}
Fig.~\ref{Fig2}(f) shows the perfect matchings of the graph in Fig.~\ref{Fig2}(c) that correspond to the individual terms in $\ket{\Lambda}$.

Finally, the four-dimensional local mode space of node $C$ is recoded as two qubits, $C_1$ and $C_2$. This recoding is a relabelling of the local basis at $C$, not an additional physical operation. With this encoding,  along with a relabelling of the local modes at the nodes, the three-node state on $B$, $C$, and $D$ can be represented as the target state $\ket{\lambda}$ in Fig.~\ref{Fig2}(d): 
\begin{equation}\label{eq:lambda_main}
\begin{aligned}
\ket{\lambda}&=\ket{\Phi^+}_{BC_1}\otimes\ket{\Phi^+}_{C_2D}
\\&=\frac{1}{2}(\ket{1111}+\ket{1122}+\ket{2211}+\ket{2222})_{BC_1C_2D}.
\end{aligned}
\end{equation}
This representation makes explicit the factorization into two Bell pairs and hence the target subgraph structure contained in $\ket{\Lambda}$. Detailed procedures are provided in the Supplementary Materials, Section~\ref{Scheme}. In Fig.~\ref{Fig2}(d), the two edges are drawn as solid lines to indicate that, after successful postselection and recoding, the corresponding links are present in the target state $\ket{\lambda}$. This representation differs from the dashed edges in Figs.~\ref{Fig2}(a)--\ref{Fig2}(c), which denote probabilistic processes associated with the random-network construction.

\begin{figure*}[!t]
\centering
\includegraphics[width=1\textwidth]{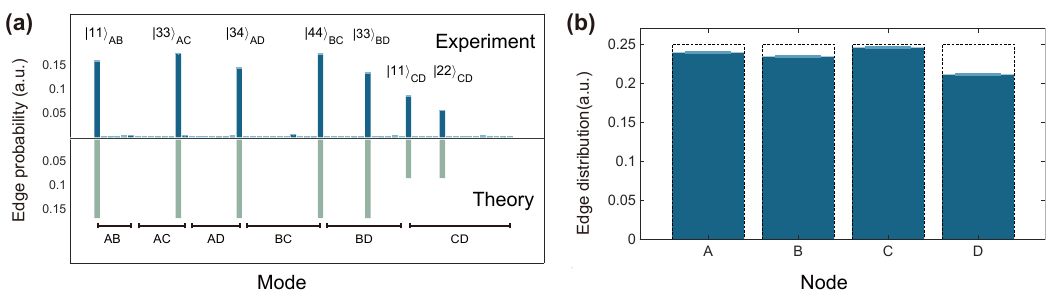}
\begin{center}
\caption{\textbf{Experimental edge probability and distribution of the resource state $\ket{G}$.} \textbf{(a),} Normalized two-fold coincidence probabilities measured in the computational basis. The results (blue bars) are shown for all possible basis combinations between node pairs, where the horizontal-axis labels denote the basis states $\ket{ij}$ with mode indices $i,j$ taking values in $\{1,3\}$ for node A, $\{1,3,4\}$ for node B, and $\{1,2,3,4\}$ for nodes C and D. The statistical overlap between the experimental and ideal distributions is $\gamma=0.9620(2)$. \textbf{(b),} Connection-probability distribution for each node (blue bars), obtained by summing the corresponding nonzero contributions in \textbf{(a)} and normalizing over all two-fold coincidence events. The overlap between experiment and theory is $\gamma=0.9636(4)$. All uncertainties are calculated from Poissonian statistics using standard error propagation}\label{Fig4n}
\end{center}
\end{figure*}
\textit{Implementation --} Fig.~\ref{Fig3} shows the integrated silicon photonic chip used to implement an optical scheme identified by the computer algorithm \textit{Melvin} \cite{PhysRevLett.116.090405} for generating the resource state $\ket{G}$ (see Supplementary Materials Section~\ref{SM_Chip} for fabrication and device details). Eight dual Mach-Zehnder interferometer microring sources (DMZI-Rs, labeled S1--S8; see Supplementary Materials Section~\ref{SM_DMZI} for details) generate photon pairs that establish the pair-creation connectivity shown in Fig.~\ref{Fig2}(b). Through spontaneous four-wave mixing (SFWM), degenerate photon pairs at 1550.81~nm are generated using dual two-color pulsed pumps \cite{Fang:13,Silverstone2014a,Zhang:19,Feng2019,Paesani2019,Chen2023} at 1544.78~nm and 1556.84~nm, each with a bandwidth of about 40~GHz. Photon pairs from sources S5–S8 are routed directly to the network nodes. For S1–S2 and S3–S4, reversed Hong–Ou–Mandel interference splits identical photon pairs into two waveguides before routing. The photons then propagate through calibrated delay-line arrays that establish node-to-node connections while preserving temporal matching and hence quantum coherence.

By encoding path modes at the four nodes, we prepare the resource state $\ket{G}$ at the white dashed line in Fig.~\ref{Fig3}. Local transformations $P_i$ with $i\in\{A,B,D\}$, defined in Eqs.~\eqref{proj1}--\eqref{proj3}, then convert $\ket{G}$ into $\ket{\Lambda}$ after postselection at the green dashed line. Local unitaries $U_i$ with $i\in\{B,C,D\}$ are applied for state analysis\cite{Clements:16}. Both $P_i$ and $U_i$ are realized using trench-isolated MZIs and phase shifters, which reduce thermal crosstalk and lower the phase-tuning power to $1.2~\mathrm{mW}$, compared with approximately $30~\mathrm{mW}$ for phase shifters without trenches. The photons are detected using superconducting nanowire single-photon detectors, and electrical signals from the detectors are processed by an FPGA-based time tagger. A four-fold coincidence event across nodes A-D is registered when one photon is detected in each of the four output paths, represented by the red, orange, green, and blue spheres in Fig.~\ref{Fig3}. Such events correspond to configurations in which each node participates in exactly one connection. At a pump power of 15~mW, we obtain a four-fold coincidence rate of approximately 0.07~Hz.


\begin{figure*}[!t]
\centering
\includegraphics[width=1\textwidth]{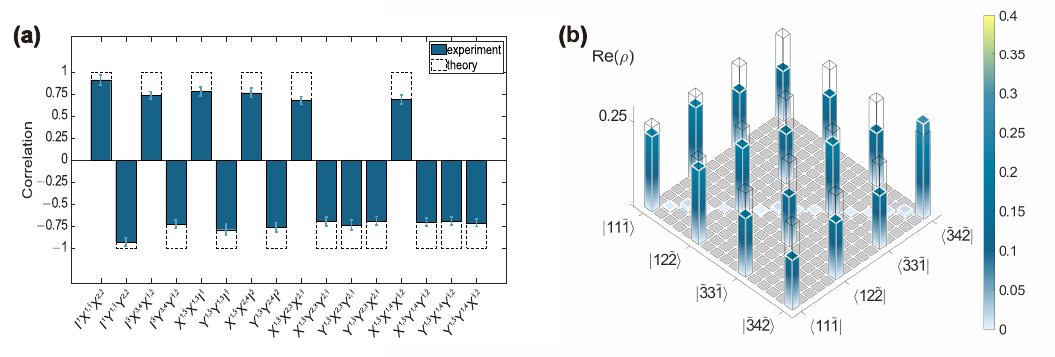}
\begin{center}
\caption{\textbf{Witnessing genuine $(2,4,2)$ high-dimensional entanglement.} \textbf{(a),} Measured expectation values for the Pauli-operator combinations used in the entanglement witness. \textbf{(b),} Witness-relevant real density-matrix elements extracted from the measurement data, together with the corresponding ideal elements of the target state $\ket{\Lambda}$. Grey elements denote unmeasured or zero components. From the measured nonzero elements, we obtain a fidelity of $0.7801(72)$, exceeding the genuine $(2,4,2)$ entanglement bound of $0.75$ by about $4.2\sigma$ (uncertainty denotes $\pm1\sigma$, obtained from 1000 Monte Carlo simulations with Poissonian error).}\label{Fig5n}
\end{center}
\end{figure*}
\textit{Experimental Results -- }
Fig.~\ref{Fig4n}(a) compares the experimental and ideal normalized two-fold coincidence distributions measured in the computational basis for the pair-creation graph used to prepare the resource state $\ket{G}$ (Fig.~\ref{Fig2}(b)). The seven nonzero bars in the ideal distribution correspond to the seven retained pair-creation edges in Fig.~\ref{Fig2}(b).  In particular, the two $CD$ components, $\ket{11}_{CD}$ and $\ket{22}_{CD}$ have half the height of the other edges. The remaining near-zero bars correspond to basis combinations absent from the resource graph. These measurements verify the intended distribution of the pair-creation graph. Coherence between the relevant four-photon alternatives is supported by the photon-indistinguishability measurements in Supplementary Materials Section~\ref{rhom&hhom} and by the entanglement witness below.  Fig.~\ref{Fig4n}(b) shows, for each node, the relative probability of forming a connection with any other node, obtained from the same two-fold data and normalized over all two-fold coincidence events. The measured distribution agrees with the ideal node-connection distribution, with small deviations attributed mainly to unbalanced on-chip losses and detection-efficiency differences among output ports.


To quantify the agreement between theory and experiment, we calculate the classical statistical overlap $\gamma = \sum_i \sqrt{p_i q_i}$, where $p_i$ and $q_i$ denote the experimental and ideal two-fold coincidence probabilities, respectively. We obtain $\gamma = 0.9620(2)$ for Fig.~\ref{Fig4n}(a) and $\gamma = 0.9636(4)$ for Fig.~\ref{Fig4n}(b). The small residual deviations are attributed to unbalanced on-chip losses and differences in detection efficiencies among the output ports.


To verify the structure required for the local recoding from $\ket{\Lambda}$ to $\ket{\lambda}$, we certify the entanglement dimensionality of the postselected state on nodes $B$, $C$, and $D$. The ideal state has a $(2,4,2)$ dimensional structure across these nodes. We therefore perform a dimension-witness measurement \cite{Erhard2018,Krenn2014a,Malik2016,Hu2020}, which rules out lower-dimensional structures, such as $(2,3,2)$, when the fidelity with the target state exceeds $0.75$ \cite{Huber2013}.

In Fig.~\ref{Fig5n}(a), we show the 16 combinations of Pauli operators used as measurement bases. For each basis, we record 16 four-fold coincidence patterns; together with an additional 16 from the computational basis, this yields 272 four-fold coincidence measurements in total. The Pauli-basis measurements probe coherence in the corresponding subspaces. From these data, we extract the 16 computational-basis populations and the 6 unique off-diagonal real parts needed for the fidelity witness. We obtain a fidelity of $F_{\textrm{exp}}=0.7801(72)$ with the ideal $(2,4,2)$ state, where the uncertainty is estimated from 1000 Monte Carlo samples assuming Poissonian photon-counting errors. Since the fidelity exceeds the bound for genuine $(2,4,2)$ entanglement, the experimental state is consistent with the target entanglement structure.



\textit{Conclusions --} We have implemented a finite photonic quantum subgraph predicted by quantum random-network theory \cite{Perseguers2010}. An integrated silicon photonic chip prepares a pair-creation resource graph corresponding to $\ket{G}$, and local mode transformations followed by postselection produce the state $\ket{\Lambda}$. By locally recoding the four-dimensional mode space of node $C$ into two logical qubits, this state can be represented as the target $\lambda$ state. Through rigorous entanglement verification, we measure a fidelity of $0.7801(72)$, exceeding the bound for genuine $(2,4,2)$ high-dimensional multipartite entanglement and confirming that the experimental state carries the entanglement structure required for the target subgraph state. The experiment does not measure the large-$N$ threshold scaling directly; rather, it realizes the finite local conversion step used in the theoretical construction.

Our silicon-photonic platform provides a scalable and reconfigurable route toward larger quantum networks: even as the network size increases and the connection probability decreases, quantum coherence continues to enable entanglement distribution between arbitrary node pairs, a capability unattainable in classical random networks at equivalent probability levels \cite{Perseguers2010}. Our results highlight that quantum entanglement enables connectivity structures beyond classical accessibility and provide an experimental support for investigating network dynamics in complex quantum networks.

\textit{Data availability.---}
The data that support the findings of this study are openly available at \url{https://github.com/NJU-Malab/Quantum-Random-Network}.

\textit{Acknowledgments --} This research was supported by HFNL Self-Deployed Project(ZB2026020200), the Natural Science Foundation of Jiangsu Province (Grants Nos. BK20240006, BK20233001), Quantum Science and Technology-National Science and Technology Major Project (Grants Nos. 2021ZD0300700 and 2021ZD0301500), Fundamental and Interdisciplinary Disciplines Breakthrough Plan of the Ministry of Education of China (JYB2025XDXM112), and Nanjing University-China Mobile Communications Group Co.,Ltd. Joint Institute.

\bibliography{reference}
\setcounter{equation}{0}
\renewcommand{\theequation}{S.\arabic{equation}}
\clearpage
\begin{widetext}
\renewcommand{\thefigure}{S\arabic{figure}} 
\setcounter{figure}{0} 

\begin{center}
 \large Supplementary Materials for “Photonic realization of a subgraph extraction in a quantum random network”
\end{center}

\subsection{Extraction of the quantum subgraphs from the quantum random network}\label{Scheme}
In classical random graph theory, the emergence of a specific subgraph exhibits a sharp threshold phenomenon: as the edge probability increases, the likelihood of finding that subgraph transitions from near zero to near one over a narrow critical region. This critical probability $p_c=cN^z$ depends on the subgraph's structure, characterized by a parameter $z$ with $z = -n/l$ and $c$ is a constant independent of $N$. Quantum random networks, however, display a fundamentally different behavior---the critical probability for generating highly connected subgraphs collapses into the $z=-2$ regime due to the coherent nature of quantum operations, a key theoretical prediction illustrated in Table.~\ref{tab:thresholds}.

Let us start from a large-scale quantum network. Initially, each pair of nodes is connected by a weakly entangled photon pair on the path-mode basis:
\begin{equation}
   \psi = \sqrt{1-\frac{p}{2}}|00\rangle + \sqrt{\frac{p}{2}}|11\rangle \quad (0 < p < 1). 
\end{equation}
Entanglement distillation is first applied to convert these links into maximally entangled Bell states $|\Phi^+\rangle$, which succeed with probability $p$ uniformly between any two nodes. Fig.~\ref{Fig2}(a) illustrates a four-node subset of this quantum random network. Here, each green dashed edge represents a probabilistic source that generates a $|\Phi^+\rangle$ Bell pair with probability $p$. The target $\lambda$ graph in Fig.~\ref{Fig2}(d) would be obtained when the two required Bell links, $B C_1$ and $C_2 D$, are simultaneously present. Since each link is generated with probability $p$, this direct formation occurs with probability $p^2$.

We now describe another theoretical conversion from a 4-node quantum network to the target $\lambda$ graph(Fig.~\ref{Fig2}(a)-(d)). In Fig.~\ref{Fig2}(a), each edge denotes a probabilistic Bell link. In the path-mode basis, such a Bell link can be written as a coherent superposition of two pair-creation components. In the graph representation, these two components are represented by two pair-creation edges, each of which creates one photon in each of the two connected nodes.

The graph in Fig.~\ref{Fig2}(b) is obtained by keeping only the pair-creation components that are relevant for the target four-photon events. For the links $AB$, $AC$, $AD$, $BC$, and $BD$, only one of the two Bell-state components is retained. These retained components are assigned the local mode labels
\begin{equation}
AB: \ket{11}_{AB},\quad
AC: \ket{33}_{AC},\quad
AD: \ket{34}_{AD},\quad
BC: \ket{44}_{BC},\quad
BD: \ket{33}_{BD}.
\end{equation}
For the link $CD$, two components are retained, corresponding to the two pair-creation edges
\begin{equation}
CD: \ket{11}_{CD},\quad \ket{22}_{CD}.
\end{equation}
The two $CD$ edges have weight $1/\sqrt{2}$, whereas all other retained edges have unit weight.

The repeated use of the same local mode label for different edges on the same node is intentional. Mode labels are local degrees of freedom at each node, and different generation alternatives can be made indistinguishable at a node by aligning them to the same local path mode. For example, the two edges $AC$ and $AD$ both occupy mode $3$ at node $A$. This mode identification removes the which-link information at node $A$ for the corresponding alternatives. In the one-photon-per-node sector, the remaining alternatives are therefore combined coherently rather than treated as distinguishable classical events.

Conditioning on one photon at each node selects the perfect matchings of the pair-creation graph. The matching $AB+CD$ gives the two terms $\ket{1111}$ and $\ket{1122}$, while the matchings $AC+BD$ and $AD+BC$ give $\ket{3333}$ and $\ket{3444}$, respectively. Taking into account the relative edge weights and normalizing the state, we obtain
\begin{equation}\label{eq:G_SI}
\ket{G}_{ABCD}=
\frac{1}{\sqrt{6}}\ket{1111}
+
\frac{1}{\sqrt{6}}\ket{1122}
+
\frac{1}{\sqrt{3}}\ket{3333}
+
\frac{1}{\sqrt{3}}\ket{3444}.
\end{equation}
Here the four indices label the local path modes at nodes $A$, $B$, $C$, and $D$, respectively.

In the experiment, we do not physically implement the full conversion from the random-network module in Fig.~\ref{Fig2}(a) to the pair-creation graph in Fig.~\ref{Fig2}(b). Instead, we directly prepare the graph in Fig.~\ref{Fig2}(b). This graph should therefore be understood as a postselected and mode-identified pair-creation representation of the four-node network module, retaining only the components needed to generate the target four-photon resource state.

Starting from $\ket{G}$, we apply local transformations $P_A$, $P_B$, $P_D$ (grey circle in Fig.~\ref{Fig2}(b)) to nodes A, B, and D, respectively:
\begin{align}
    |\tilde{1}\rangle_{A} &= \frac{1}{\sqrt{2}}(|1\rangle_{A}+|3\rangle_{A}), \quad |\tilde{3}\rangle_{A} = \frac{1}{\sqrt{2}}(|1\rangle_{A}-|3\rangle_{A}), \label{proj1} \\
    |\tilde{3}\rangle_{B} &= \frac{1}{\sqrt{2}}(|3\rangle_{B}+|4\rangle_{B}), \quad |\tilde{4}\rangle_{B} = \frac{1}{\sqrt{2}}(|3\rangle_{B}-|4\rangle_{B}), \label{proj2} \\
    |\tilde{1}\rangle_{D} &= \frac{1}{\sqrt{2}}(|1\rangle_{D}+|3\rangle_{D}), \quad |\tilde{2}\rangle_{D} = \frac{1}{\sqrt{2}}(|2\rangle_{D}+|4\rangle_{D}), \notag \\
    |\tilde{3}\rangle_{D} &= \frac{1}{\sqrt{2}}(|1\rangle_{D}-|3\rangle_{D}), \quad |\tilde{4}\rangle_{D} = \frac{1}{\sqrt{2}}(|2\rangle_{D}-|4\rangle_{D}). \label{proj3}
\end{align}
Followed by postselection on specific output modes, it yields the state $|\Lambda\rangle$ (Fig.~\ref{Fig2}(c)) with fixed success probability 1/6.
\begin{equation}\label{eq:Lambda_SI}
\begin{aligned}
\ket{\Lambda}_{ABCD} &=\frac{1}{2}(\ket{\tilde{1}11\tilde{1}}+\ket{\tilde{1}12\tilde{2}}+\ket{\tilde{1}\tilde{3}3\tilde{1}}+\ket{\tilde{1}\tilde{3}4\tilde{2}})_{ABCD}\\
&=\frac{1}{2}\ket{\tilde{1}}_A(\ket{11\tilde{1}}+\ket{12\tilde{2}}+\ket{\tilde{3}3\tilde{1}}+\ket{\tilde{3}4\tilde{2}})_{BCD},
\end{aligned}
\end{equation}
where the details of the path modes are shown in Fig.~\ref{Fig2}(f). By triggering the photon in node A , we generate an entangled state $|\Lambda\rangle_{BCD}$ with Schmidt-rank vector $SRV_{b|cd,c|bd,d|bc}=(2,4,2)$, which represents the ordered ranks of the single photon reductions of the state density operator\cite{Huber2013} (each number is the dimension of each particle). By relabeling the path basis states as:
\begin{align}\label{eq:relabel}
&\ket{\tilde{3}}_{B} \mapsto \ket{2}_{B},\quad
\ket{1}_{C} \mapsto \ket{11}_{C_1C_2},\quad
\ket{2}_{C} \mapsto \ket{12}_{C_1C_2},\nonumber \quad\\  
&\ket{3}_{C} \mapsto \ket{21}_{C_1C_2},\quad
\ket{4}_{C} \mapsto \ket{22}_{C_1C_2},\quad
\ket{\tilde{1}}_{D} \mapsto \ket{1}_{D},\quad \nonumber\\
&\ket{\tilde{2}}_{D} \mapsto \ket{2}_{D},
\end{align}
we transform the network state of nodes B,C,D into the final subgraph state $\ket{\lambda}$(Fig.~\ref{Fig2}(d)): 
\begin{equation}\label{eq:lambda_SI}
\begin{aligned}
\ket{\lambda}&=\ket{\phi}^{+}_{BC_1}\otimes\ket{\phi}^{+}_{C_2D}
\\&=\frac{1}{2}(\ket{1111}+\ket{1122}+\ket{2211}+\ket{2222})_{BC_1C_2D},
\end{aligned}
\end{equation}
Experimentally, we verify the state \(|\Lambda\rangle\) on nodes \(B, C, D\). The target \(\lambda\) graph is then obtained via a local basis recoding of node \(C\) into two logical qubits \(C_1\) and \(C_2\). This operation preserves the underlying entanglement structure, resulting in a state that factorizes into two Bell pairs (Fig.~\ref{Fig2}(d)). This demonstrates the conversion of a multipartite entangled resource into a structured graph with the targeted high-degree connections.

We have thus introduced a procedure for realizing a \(\lambda\)-quantum subgraph from a subset of a 4-node quantum random network by leveraging the coherence of photonic paths within the network. While this process occurs with a single-shot probability of \(p^2\) that mirrors classical graph behavior, the underlying structure is composed of subgraph configurations residing in the \(z=-2\) regime (Table.~\ref{tab:thresholds}). Consequently, this leads to a lower critical probability threshold in the limit of an infinite quantum random network\cite{Perseguers2010}.

\subsection{Details of integrated silicon-photonic components}\label{SM_Chip}
In our experimental work, we couple a dual-wavelength pulsed pump laser\cite{Chen2023} to the chip's planar waveguide through vertical coupling. The instrumentation designed for accurate optical coupling alignment in fiber-chip array interconnects is shown in Fig.~\ref{Cou1}. The blue and orange square is edge couplers (EC) and grating couplers (GC), respectively. Through the optimization of optical coupling for the GC1-GC4 configuration, we ensured that the vertical coupling achieved the optimal coupling position. By simultaneously optimizing the EC1-EC2 and EC3-EC4 configurations, we guaranteed that the edge coupling reached its optimal alignment position.
\begin{figure}[htp!]
\centering
\includegraphics[width=1\linewidth]{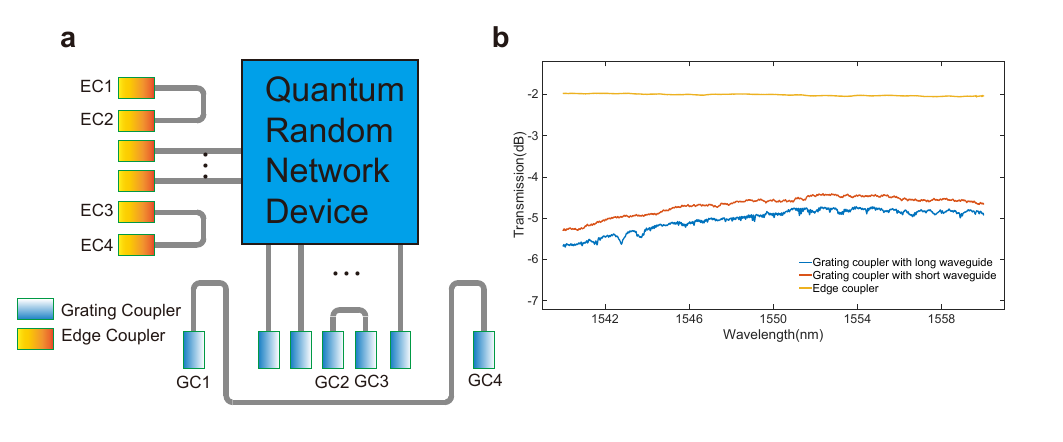}
\begin{center}
\caption{\textbf{a,} The instrumentation designed for accurate optical coupling alignment in fiber-chip array interconnects. The blue and orange square is Edger Couplers and Grating Couplers, respectively. Through the optimization of optical coupling for the GC1-GC4 configuration, we ensured that the vertical coupling achieved the optimal coupling position. By simultaneously optimizing the EC1-EC2 and EC3-EC4 configurations, we guaranteed that the edge coupling reached its optimal alignment position. \textbf{b,} The spectral transmission of the edge coupler EC1-EC2 (blue line), the short grating coupler pair GC2-GC3 (red line) and the long grating coupler pair GC1-GC4 (blue line). A comparative analysis of short and long grating coupler transmission spectra is conducted to quantitatively estimate the waveguide propagation losses. }\label{Cou1}
\end{center}
\end{figure}
The vertical coupling loss is measured to be around $4- 5~\text{dB/facet}$ for pump wavelengths of $1544.78~\text{nm}$ and $1556.84~\text{nm}$, respectively. The waveguide loss is estimated to be approximately $2~\text{dB/cm}-2.5~\text{dB/cm}$, estimated by utilizing the length difference between GC1-GC4 and GC2-GC3. The loss of a single waveguide crossing being about $0.2-0.3~\text{dB}$ and the loss of each multimode interference (MMI) coupler is about $0.5-0.6~\text{dB}$, according to the typical performance in our experiment\cite{Chen2023}. 

In the experiment, we employed deep etching technology to thermally isolate the thermal phase modulators in all Mach-Zehnder interferometers (MZIs). This reduced the periodic modulation power of the modulators to $1.2~\text{mW}$, compared to approximately $30~\text{mW}$ for untreated phase shifters. This modification ensured that the phase shifters in all unitary matrices introduced much less thermal crosstalk strongly moving the resonance peaks of the dual Mach-Zehnder interferometers resonators.

At nodes A, B, C, and D, the dimension of photon path modes were 2, 3, 4, and 4, respectively, resulting in different numbers of devices and varying waveguide lengths for routing and modulation.
 After photon generation and subsequent operations, photons were coupled into the fiber array using edge couplers. With refractive‑index‑matching fluid (n = 1.43) applied, the loss per edge coupler is about 2.3 dB per facet. On different output channels, we incorporated cascaded dense wavelength division multiplexing (DWDM) filters, with the maximum loss variation among the eight-channel cascaded filter sets being about $0.88~\text{dB}$.  

\subsection{Characterization of the integrated dual Mach-Zehnder interferometer micro-ring (DMZI-R) source}\label{SM_DMZI}

We realize a quadruple-coupler ring resonator using two pulley waveguides wrapped around the ring’s input and output sides, as illustrated in  Fig.~\ref{SM_DMZI_Pri}. The waveguide and ring resonator configuration effectively forms two distinct asymmetric Mach-Zehnder interferometers ($\mathrm{AMZI}_1$ and $\mathrm{AMZI}_2$). Within this setup, $\mathrm{AMZI}_1$ is defined by two directional couplers(with the same coupling coefficient $\kappa_1$) along with two arms $r_1$ and $r3$, while $\mathrm{AMZI}_2$ comprises two directional couplers(with the same coupling coefficient $\kappa_2$) and two arms $r_2$ and $r_5$. The ring resonator possesses a radius denoted as 15 $\mu m$. Additional waveguide segments, $r_4$ and $r_6$, serve as the inter-coupler connections. We assume the directional couplers introduce negligible physical length and exhibit a constant coupling ratio across the operational wavelength band. Crucially, with carefully designed waveguide lengths, the free spectral range (FSR) of each AMZI can be made twice that of the ring resonator.
\begin{figure}[htp!]
\centering
\includegraphics[width=0.5\linewidth]{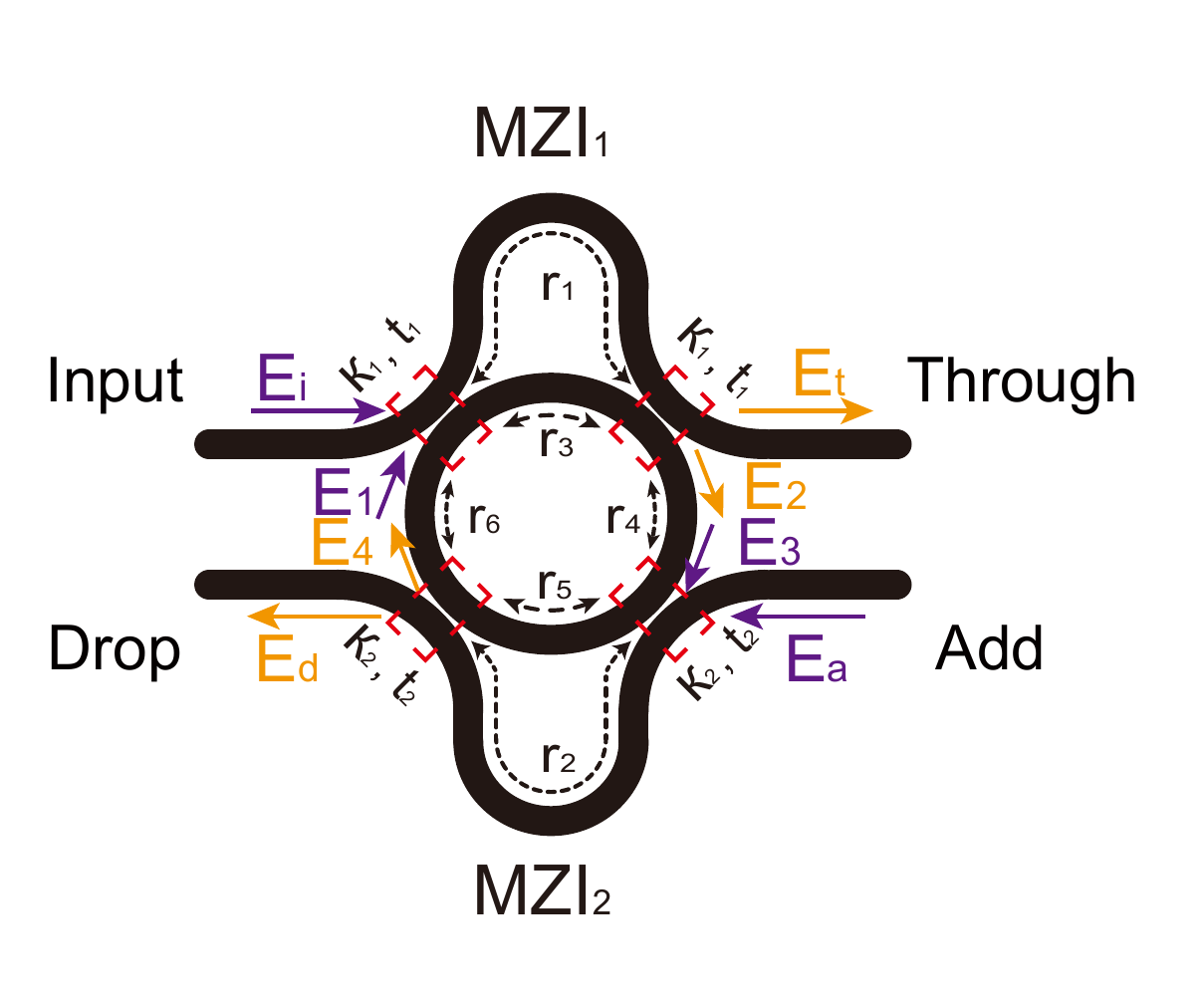}
\begin{center}
\caption{The schematic of a DMZI-R. A DMZI-R can be divided into part I, part II and part III. Part I and part III is mainly formed by a AMZI and two straight waveguides, while part II is formed by two bend waveguides connecting part I and part III.}\label{SM_DMZI_Pri}
\end{center}
\end{figure}
By assuming no electric field input at $E_a$, the transmittance characteristics are derived as\cite{Lu2020}:
\begin{equation}
    E_{d}=\frac{\gamma^{4}t_{1} \kappa_{1}t_{2}\kappa_{2}\left( \varphi_{1}+ \varphi_{3}\right) \varphi_{4}\left( \varphi_{2}+ \varphi_{5}\right)}{1-\gamma^{4}\left(t_{2}^2 \varphi_{5}-\kappa_{2}^2 \varphi_{2}\right) \varphi_{6}\left(t_{1}^2 \varphi_{3}-\kappa_{1}^2 \varphi_{1}\right) \varphi_{4}},\label{eq:SM12}
\end{equation}
and
\begin{equation}
    E_{t}=\frac{\gamma^{6}\left(t_{2}^2 \varphi_{1} \varphi_{3} \varphi_{4} \varphi_{5} \varphi_{6}-\kappa_{2}^2 \varphi_{1} \varphi_{2} \varphi_{3} \varphi_{4} \varphi_{6}\right)-\gamma^{2}\left(t_{1}^2 \varphi_{1}-\kappa_{1}^2 \varphi_{3}\right)}{\gamma^{4}\left(t_{2}^2 \varphi_{5}-\kappa_{2}^2 \varphi_{2}\right) \varphi_{6}\left(t_{1}^2 \varphi_{3}-\kappa_{1}^2 \varphi_{1}\right) \varphi_{4}-1},\label{eq:SM13}
\end{equation}
where $\varphi_i=e^{-\alpha r_i}e^{i\theta_i}$, $\theta_i=\beta r_i$($i\in \{1, ..., 6\}$, with $\alpha$ being the attenuation constant and $\beta$ being the propagation constant). Eq.~\eqref{eq:SM12} and Eq.~\eqref{eq:SM13} indicate that the DMZI-R's transmission characteristics depends critically on phase term $\varphi_i$. Specifically, we investigate a design where the four phase terms for the ring are equal ($\varphi_{3}=\varphi_{4}=\varphi_{5}=\varphi_{6}=\varphi_{r}=\exp [-\alpha 2 \pi r / 4] \cdot \exp [i \beta 2 \pi r / 4]$). The resonator features four distinct closed loops; the imposed resonance conditions governing their round-trip phase shifts are
\begin{equation}
    \begin{aligned} \theta_{1}+\theta_{2}+2 \theta_{r} & =N_{1} \times \pi \\ \theta_{1}+3 \theta_{r} & =N_{2} \times \pi \\ \theta_{2}+3 \theta_{r} & =N_{3} \times \pi \\ 4 \theta_{r} & =N_{4} \times \pi ,\end{aligned}\label{eq:confineeq}
\end{equation}
where $N_i\in Z (i=1,2,3,4)$. Eq.~\eqref{eq:confineeq} gives some restriction on the relationship between $\theta_{1/2}$ and $\theta_r$. By selecting the phase conditions $\phi_1 = \theta_1 - \theta_r = 2N\pi$ and $\phi_2 = \theta_2 - \theta_r = (2N+1)\pi$, the pump leakage at the through port is minimized while the pump light is confined within the resonator and the generated photons are directed to the drop port. The input-through and input-drop transmission spectrum of a typical DMZI-R are summarized in Fig.~\ref{Cef1}. The nominally designed $\mathrm{AMZI_1}$($\mathrm{AMZI_2}$) arm length difference is $47.8~\mu m$($48~\mu m$). A radius of $15~ \mu m$ is designed for the ring. All DMZI-Rs demonstrate noise suppression exceeding 20 dB. The quality factor ($Q$) values of the eight integrated DMZI-Rs on the chip ranged from $1.806 \times 10^4$ to $6.536 \times 10^4$.

\begin{figure}[htp!]
\centering
\includegraphics[width=180mm]{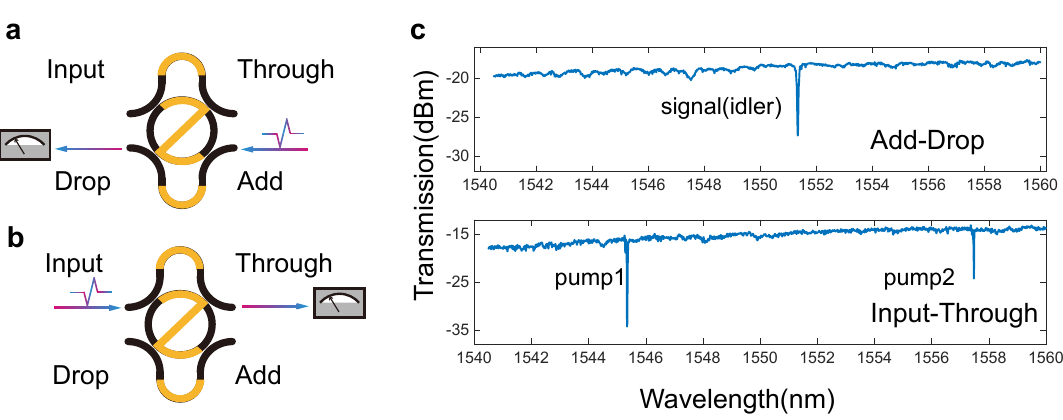}
\begin{center}
\caption{\textbf{a,} Schematic diagram showing add-to-drop spectral transmission characterization of a DMZI-R with continuous-wave laser. \textbf{b,} Schematic diagram showing input-to-through spectral transmission characterization of a DMZI-R with continuous-wave laser. \textbf{c,} The transmission spectrum for input-to-drop (up) and add-to-drop (bottom).}\label{Cef1}
\end{center}
\end{figure}

\subsection{Characterisations of indistinguishability of the photons in network}\label{rhom&hhom}
To create and manipulate entanglement between nodes, the photons in network are required to interfere perfectly, demanding the indistinguishability between photons. The factors that degrade on‑chip indistinguishability were analyzed and characterized in previous work\cite{Lu2020,Chen2023}. In our experiment, we perform reversed Hong-Ou-Mandel (RHOM) and heralded Hong-Ou-Mandel (HHOM) quantum interference to characterize the spectrum and purity influence between DMZI-Rs. These experiments are typical for a assessment for the indistinguishability between photons.
\begin{figure}[htp!]
\centering
\includegraphics[width=1\linewidth]{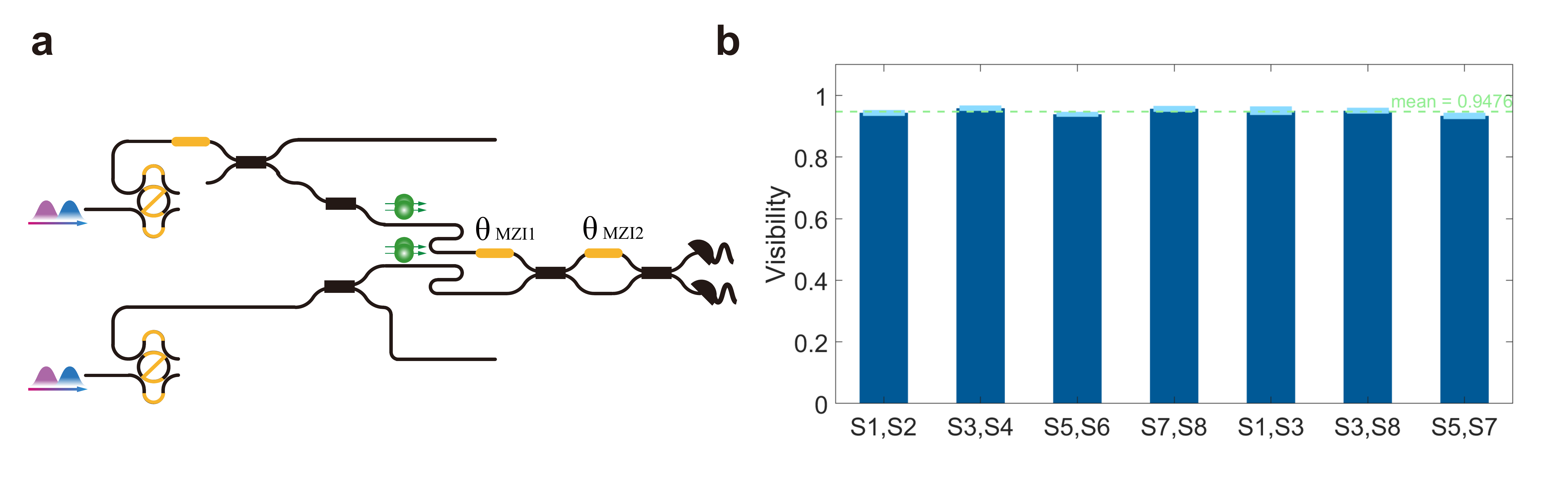}
\begin{center}
\caption{\textbf{a,} Schematic of a RHOM quantum interference. By varying the phase shift \(\theta_{\mathrm{MZI1}}\) and setting \(\theta_{\mathrm{MZI2}} = \pi/2\), one can observe the RHOM interference pattern.  
\textbf{b,} Representative RHOM quantum interference patterns obtained by varying \(\theta_{\mathrm{MZI2}}\) across the eight DMZI-Rs. The average fringe contrast across all measured patterns is 0.9476, as determined by Eq.~\eqref{rhom_vis}. These results characterize the high degree of spectral overlap across DMZI-Rs S1--S8. All uncertainties are derived
from Poissonian statistics and standard error propagation.}\label{rhom}
\end{center}
\end{figure}
The spectrum overlap of photons generated from two different DMZI-Rs can be characterized by the reversed Hong-Ou-Mandel quantum interference, the schematic is shown in Fig.~\ref{rhom}a. The contrast of a RHOM pattern is defined as:
\begin{equation}\label{rhom_vis}
    C=\mathrm{\frac{CC_{max}-CC_{min}}{CC_{max}+CC_{min}}}.
\end{equation}
Here $\mathrm{CC_{max}}$ ($\mathrm{CC_{min}}$) denotes the maximum (minimum) two-fold coincidence counts in the RHOM quantum interference pattern. When photons generated from two DMZI-Rs exhibit perfect spectral overlapping and probability amplitude matching, the pattern contrast reaches $1$. This two-photon process can be extended to a probabilistic splitting scheme involving probabilistic photon-pair sources, as shown in Fig.~\ref{rhom}a. In this scheme, photon pairs undergo probabilistic splitting and still exhibit RHOM interference upon reaching the two-dimensional analysis matrix (Mach-Zehnder interferometer with $\theta_{\mathrm{MZI1}}$ and $\theta_{\mathrm{MZI2}}$). The amplitude mismatch of photon pairs is eliminated by adjusting the phase shifter in the pump distributor. As shown in Fig.~\ref{rhom}b, the average fringe contrast across all measured patterns is 0.947, as determined by Eq.~\eqref{rhom_vis}. These results characterize the high degree of spectral overlap across DMZI-Rs S1--S8. All uncertainties are derived from Poissonian statistics and standard error propagation.
\begin{figure}[htp!]
\centering
\includegraphics[width=1\linewidth]
{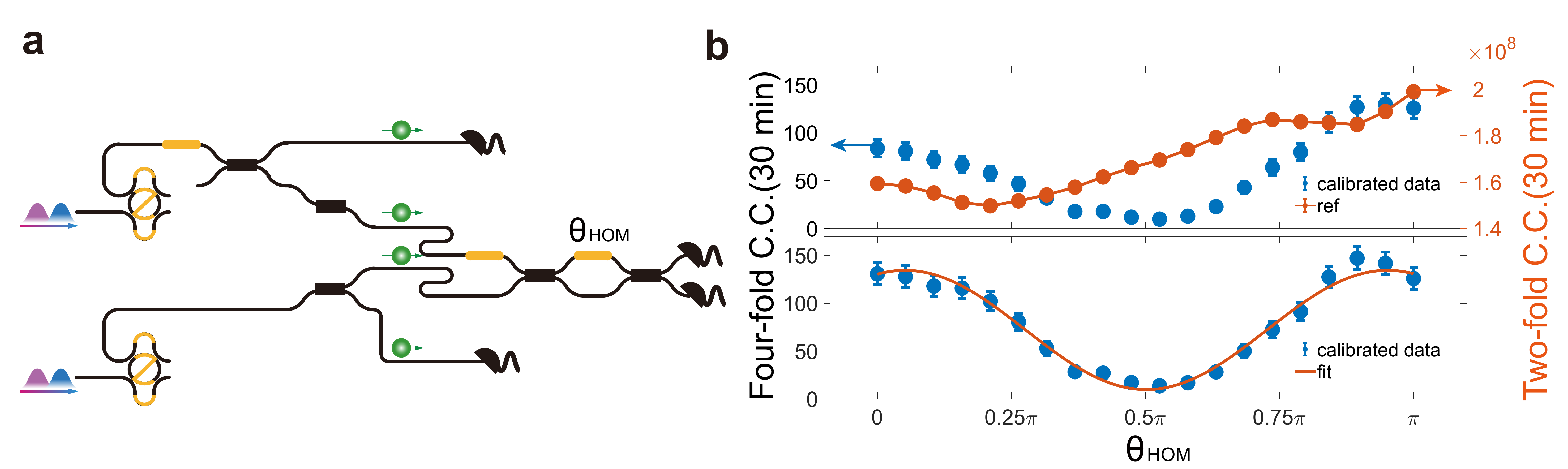}
\begin{center}
\caption{\textbf{a,} Schematic of the heralded Hong-Ou-Mandel (HHOM) quantum interference setup. Varying the phase shift \(\theta_{\mathrm{MZI2}}\) enables the HHOM quantum interference measurement. \textbf{b,} Experimental HHOM interference patterns. \textit{Upper panel}: Raw experimental data (blue circles) and the reference signal from an isolated DMZI-R (red circles), used to monitor power and coupling efficiency drift. \textit{Lower panel}: Calibrated experimental data (blue circles) obtained using Eq.~\eqref{cal_raw}, overlaid with the fitting curve (red line). The pattern visibilities, calculated using Eq.~\eqref{contrast}, are \(0.7872(636)\) (raw) and \(0.8289(441)\) (calibrated). Each data point represents a four-fold coincidence count integrated over 30 minutes. All uncertainties are derived from Poissonian statistics and standard error propagation.}\label{hhom}
\end{center}
\end{figure}

Photon indistinguishability is further characterized using heralded Hong-Ou-Mandel (HHOM) quantum interference. The experimental schematic is shown in Fig.~\ref{hhom}a. By heralding two photons, we measure the HHOM interference pattern by varying the phase shifter $\theta_{\mathrm{HOM}}$ in the Mach-Zehnder interferometer. The visibility of the HHOM quantum interference pattern is defined as\cite{Bao2023}:
\begin{equation}
    V = \frac{\mathrm{CC_{max}} - 2\mathrm{CC_{min}}}{\mathrm{CC_{max}}}\label{contrast}
\end{equation}
here $\mathrm{CC_{max}}$ ($\mathrm{CC_{min}}$) denotes the maximum (minimum) four-fold coincidence counts. At phases $\theta_{\mathrm{HOM}} = 0$ or $\pi$, quantum interference is absent and photons distribute randomly between the output paths, maximizing the coincidence counts. When $\theta_{\mathrm{HOM}} = \pi/2$, identical photons interfere destructively, producing a characteristic Hong-Ou-Mandel dip. The HHOM interference visibility depends critically on spectral overlap and photon purity. Experimental results (Fig.~\ref{hhom}b) yield a raw visibility of $0.7872(636)$ calculated via Eq.~\eqref{contrast}. Each data point represents four-fold coincidence counts integrated over 30-minute intervals, with the complete measurement spanning 10 hours.

Pump power drift on-chip, induced by input power fluctuations and coupling efficiency variations, is monitored using a reference DMZI-R (red trace, upper subplot). We use the following equation to calibrate the four-fold coincidence counts:
\begin{equation}
    \mathrm{CC_{4F-cal}=CC_{4f-raw}\cdot\eta^2}\label{cal_raw}
\end{equation}
where $\mathrm{\eta=CC_{2f-max}/CC_{2f}}$, $\mathrm{CC_{2f}}$ is two-fold coincidence counts in 30 minutes of the reference DMZI-R and $\mathrm{CC_{2f-max}}$ the maximum two-fold coincidence counts. After drift calibration with this reference, the corrected visibility is $0.8289(441)$, which highlights the enhanced photon indistinguishability achievable with DMZI-R devices.

\subsection{Entanglement witness by fidelity}\label{EW_SM}
Full reconstruction of our (2,4,2) $\lambda$ state by standard quantum state tomography requires the measurement of all $136$ elements of density matrix, which is consuming for both computing resources and time for the multiphoton cases. In order to verify a $\lambda$ graph like entanglement type distribution between nodes, we use the method called entanglement witness developing recently. The goal of this method is to prove that the experimental state cannot be decomposed into states of a smaller dimensionality structure. Following the implementation in previous experiments \cite{Krenn2014a,Hu2020,Malik2016}, we 
witness the fidelity $F_{exp}$ between experimental state and the target state $|\Lambda\rangle$, where  
\begin{equation}
\begin{aligned}
F_{exp} &=Tr(\rho_{exp}|\Lambda\rangle\langle\Lambda|)\\
=\frac{1}{4}(&\langle11\tilde1|\rho_{exp}|11\tilde1\rangle+\langle12\tilde2|\rho_{exp}|12\tilde2\rangle\\
+&\langle\tilde33\tilde1|\rho_{exp}|\tilde33\tilde1\rangle+\langle\tilde34\tilde2|\rho_{exp}|\tilde34\tilde2\rangle\\
+&\langle11\tilde1|\rho_{exp}|12\tilde2\rangle+\langle12\tilde2|\rho_{exp}|11\tilde1\rangle\\
+&\langle\tilde33\tilde1|\rho_{exp}|\tilde34\tilde2\rangle+\langle\tilde34\tilde2|\rho_{exp}|\tilde33\tilde1\rangle\\
+&\langle11\tilde1|\rho_{exp}|\tilde33\tilde1\rangle+\langle\tilde33\tilde1|\rho_{exp}|11\tilde1\rangle\\
+&\langle12\tilde2|\rho_{exp}|\tilde34\tilde2\rangle+\langle\tilde34\tilde2|\rho_{exp}|12\tilde2\rangle\\
+&\langle11\tilde1|\rho_{exp}|\tilde34\tilde2\rangle+\langle\tilde34\tilde2|\rho_{exp}|11\tilde1\rangle\\
+&\langle12\tilde2|\rho_{exp}|\tilde33\tilde1\rangle+\langle\tilde33\tilde1|\rho_{exp}|12\tilde2\rangle)
\end{aligned}
\end{equation}
For the sake of simplicity and clarity, we leave out the tilde on the letters here. The four diagonal terms can be obtained by one single projection $\langle ijk|\rho_{exp}|ijk \rangle=C_{ijk}/C_{N}$, $i=1,\tilde3$, j=1,2,3,4, k=$\tilde1,\tilde2$. The normalization $C_{N}=\sum_{i}\sum_{j}\sum_{k}C_{ijk}$, where $C_{ijk}$ is the coincidence counts in different output combination under different projection measurement.
All the 16 projections in computational basis are measured simultaneously. Due to $\langle ijk|\rho_{exp}|lmn\rangle+\langle lmn|\rho_{exp}|ijk\rangle=2\mathrm{Re} (\langle ijk|\rho_{exp}|lmn\rangle)$, only 6 unique in our systems need to be measured ($\mathrm{Re} (\langle 11\tilde1|\rho_{exp}|12\tilde2\rangle)$, $\mathrm{Re} (\langle \tilde33\tilde1|\rho_{exp}|\tilde34\tilde2\rangle)$, $\mathrm{Re} (\langle 11\tilde1|\rho_{exp}|\tilde33\tilde1\rangle)$, $\mathrm{Re} (\langle 12\tilde2|\rho_{exp}|\tilde34\tilde2\rangle)$, $\mathrm{Re} (\langle 11\tilde1|\rho_{exp}|\tilde34\tilde2\rangle)$, $\mathrm{Re} (\langle 12\tilde2|\rho_{exp}|\tilde33\tilde1\rangle)$). For the first 4 two-partite superposition terms ($\mathrm{Re} (\langle 11\tilde1|\rho_{exp}|12\tilde2\rangle)$, $\mathrm{Re} (\langle \tilde33\tilde1|\rho_{exp}|\tilde34\tilde2\rangle)$, $\mathrm{Re} (\langle 11\tilde1|\rho_{exp}|\tilde33\tilde1\rangle)$, $\mathrm{Re} (\langle 12\tilde2|\rho_{exp}|\tilde34\tilde2\rangle)$), we can measure them using the following decomposition:
\begin{equation}
\begin{aligned}
\mathrm{Re} (\langle ijk|\rho_{exp}|imn\rangle)
= 
\frac{N^{ijk,imn}}{4}(\langle\sigma_I^i\otimes \sigma^{jm}_{x}\otimes\sigma^{kn}_{x}\rangle-  \langle\sigma_I^i \otimes \sigma^{jm}_{y}\otimes\sigma^{kn}_{y}\rangle)
\end{aligned}\label{Puail1}
\end{equation}
where Pauli matrices $\sigma^{ab}_{x}=|a\rangle\langle b|+|b\rangle\langle a|$,  $\sigma^{ab}_{y}=i|a\rangle\langle b|-i|b\rangle\langle a|$, computational basis $\sigma_I^i=|i\rangle\langle i|$ and 
\begin{equation}
\begin{aligned}
 N^{ijk,imn}=Tr((|ijk\rangle\langle ijk|+|imn\rangle\langle imn|)\rho_{exp}(|ijk\rangle\langle ijk|+|imn\rangle\langle imn|)).
\end{aligned}
\end{equation}
Here we note that $\rho_{exp}$ is a high-dimensional quantum system, when we observe $\sigma_{x}^{jm}$ and $\sigma_{y}^{jm}$($\sigma_{x}^{kn}$ and $\sigma_{y}^{kn}$) in its subspace, the normalization $N^{il,jm,kn}$ in the experimental system can be measured simultaneously through a complete collection of the distribution in all dimensional. $N^{il,jm,kn}$ should be consistent under different measurement results, we estimate it from the distribution under computational basis. For example, for the $\langle 111|\rho_{exp}|122\rangle$, we perform $\sigma_I^1\otimes\sigma^{12}_{x}\otimes \sigma^{\tilde{1}\tilde{2}}_{x}$ and 
$\sigma_I^1\otimes\sigma^{12}_{y}\otimes \sigma^{\tilde{1}\tilde{2}}_{y}$, we get:
\begin{equation}
\begin{aligned}
 \langle\sigma_I^1\otimes\sigma^{12}_{x}\otimes \sigma^{\tilde{1}\tilde{2}}_{x}\rangle=\frac{C_{11\tilde1}+C_{12\tilde2}-C_{12\tilde1}-C_{11\tilde2}}{C_{11\tilde1}+C_{12\tilde2}+C_{12\tilde1}+C_{11\tilde2}}
\end{aligned}
\end{equation}
For the last 2 three-partite superposition terms ($\mathrm{Re} (\langle 11\tilde1|\rho_{exp}|\tilde34\tilde2\rangle)$, $\mathrm{Re} (\langle 12\tilde2|\rho_{exp}|\tilde33\tilde1\rangle)$), we can measure them using the following decomposition:

\begin{equation}
\begin{aligned}
\mathrm{Re} (\langle ijk|\rho_{exp}|lmn\rangle)
= \frac{N^{ijk,lmn}}{8}(\langle\sigma^{il}_{x}\otimes\sigma^{jm}_{x} \otimes\sigma^{kn}_{x}\rangle
- \langle\sigma^{il}_{x}\otimes\sigma^{jm}_{y} \otimes\sigma^{kn}_{y}\rangle
\\- \langle\sigma^{il}_{y}\otimes\sigma^{jm}_{x} \otimes\sigma^{kn}_{y}\rangle
- \langle\sigma^{il}_{y}\otimes\sigma^{jm}_{y} \otimes\sigma^{kn}_{x}\rangle)
\end{aligned}\label{Puail2}
\end{equation}

The experimental result of all the observation combination is shown in Fig.~\ref{Fig5n}a. The confidence intervals in Fig.~\ref{Fig5n}a is calculated by assuming the detection photon numbers under different measurement to be Poisson distribution. 

To certify the entanglement dimensionality between C and BD, we compare the experimental fidelity with the maximal fidelity achievable by states of limited Schmidt rank. For a given class of states $\mathcal{S}$, the maximum fidelity with our target state $\lvert\Lambda\rangle$ is defined as
\[
F_{\text{max}}(\mathcal{S}) = \max_{\sigma \in \mathcal{S}} \operatorname{Tr}\bigl(\sigma \lvert\Lambda\rangle\langle\Lambda\rvert\bigr).
\]

For a pure state with Schmidt rank vector $(2,3,2)$ (i.e., at most 2, 3, 2 across the three bipartitions), we denote $\mathcal{S} = (2,3,2)$ and obtain
\[
F_{\text{max}}^{(2,3,2)} = \frac{3}{4}.
\]
Exceeding this bound proves that the state cannot be described by a $(2,3,2)$ state, implying that the C--BD partition requires at least a 4‑dimensional entangled subspace (genuine four‑dimensional entanglement).

For a $(2,2,2)$ state (all bipartitions of Schmidt rank $\le 2$), $\mathcal{S} = (2,2,2)$ and
\[
F_{\text{max}}^{(2,2,2)} = \frac{1}{2}.
\]
A fidelity above $1/2$ rules out genuinely 2‑dimensional entangled state, thus establishing that the C--BD entanglement is truly three‑dimensional (or higher).

Both values follow from the Schmidt coefficients of $\lvert\Lambda\rangle$: $\{1/2,1/2\}$ for the D$\vert$BC and B$\vert$CD partitions, and $\{1/4,1/4,1/4,1/4\}$ for C$\vert$BD, using the results of Ref.~\cite{Fickler2014}.

\subsection{Entanglement witness by visibility}
In the preceding section, we presented a fidelity‑based witness to certify the entanglement dimensionality. That approach compared the measured overlap with a reference state $\text{SRV}(2,4,2)$ against the maximal possible overlap achievable by any state with Schmidt number $\le 3$ (i.e., $\text{SRV}(2,3,2)$), thereby demonstrating genuine four‑dimensional entanglement. We note that an alternative witness constructed from visibilities, evaluated using the same experimental data, yields a consistent conclusion (see the following subsection for details). The fidelity‑based result alone, however, already provides a definitive proof of four‑dimensional entanglement between $C$ and $BD$, and the visibility‑based witness is presented only as an illustration of a different theoretical construction. 

We consider a tripartite system consisting of subsystems $B$, $D$, and $C$. Here $B$ is a qubit with basis $\{|1\rangle_B, |3\rangle_B\}$, $D$ is a qubit with basis $\{|1\rangle_D, |2\rangle_D\}$, and $C$ is a four‑level system with basis $\{|1\rangle_C, |2\rangle_C, |3\rangle_C, |4\rangle_C\}$. The three‑body state of interest is expressed in terms of these basis states as
\begin{equation}
\label{eq:state}
|\psi\rangle_{BDC} = \frac{1}{2}\bigl(
|111\rangle
+ |122\rangle
+ |313\rangle
+ |324\rangle
\bigr),
\end{equation}
where the kets are ordered as $|B D C\rangle$.
The system $E$, composed of subsystems $B$ and $D$, is effectively four‑dimensional. Its orthonormal basis vectors are defined as
\begin{equation}
|1\rangle_E = |1\rangle_B|1\rangle_D,\quad
|2\rangle_E = |1\rangle_B|2\rangle_D,\quad
|3\rangle_E = |3\rangle_B|1\rangle_D,\quad
|4\rangle_E = |3\rangle_B|2\rangle_D.
\end{equation}
Thus the effective four‑dimensional system $E$ has the above basis. For a subspace spanned by $\{|a_E\rangle,|b_E\rangle\}$, we introduce Pauli‑type operators acting on system $E$ as
\begin{equation}
X_E^{ab} = |E_a\rangle\langle E_b| + |E_b\rangle\langle E_a|,\qquad
Y_E^{ab} = i|E_a\rangle\langle E_b| - i|E_b\rangle\langle E_a|,\qquad 
Z_E^{ab}=|E_a\rangle\langle E_a| - |E_b\rangle\langle E_b|,
\end{equation}
where $ab \in \{12,34,13,24,14,23\}$. For the subspaces spanned by mode pairs $\{|1\rangle_E,|2\rangle_E\}$, $\{|3\rangle_E,|4\rangle_E\}$, $\{|1\rangle_E,|3\rangle_E\}$, and $\{|2\rangle_E,|4\rangle_E\}$, the operators $X_E^{ab}$ and $Y_E^{ab}$ can be decomposed directly into local operators on $B$ and $D$. For example, for the subspace $\{|1\rangle_E,|2\rangle_E\}$ we have
\begin{equation}
X_E^{12}  = \sigma_I^{1}\otimes\sigma_x^{12},\qquad
Y_E^{12}  = \sigma_I^1\otimes\sigma_y^{12}.
\end{equation}
For the subspaces spanned by $\{|1_E\rangle,|4_E\rangle\}$ and $\{|2_E\rangle,|3_E\rangle\}$, the Pauli‑type operators are still decomposable into local operations on $B$ and $D$, albeit in a less direct manner. For instance, $X_E^{14}$ can be expressed as
\begin{equation}
X_E^{14}  = \sigma_x^{11}\otimes\sigma_x^{32} - \sigma_y^{11}\otimes\sigma_y^{32},\qquad
Y_E^{14}  = \sigma_x^{11}\otimes\sigma_y^{32} + \sigma_y^{11}\otimes\sigma_x^{32}.
\end{equation}
Three visibilities $V_x$, $V_y$, and $V_z$ are defined as the correlations between system $E$ (i.e., $BD$) and system $C$ within a two‑dimensional subspace corresponding to mode pairs $\{ij\}$:
\begin{equation}
V_x^{ab} = \langle X_E^{ab} \sigma_x^{ab}\rangle,\qquad 
V_y^{ab} = \langle Y_E^{ab} \sigma_y^{ab}\rangle,\qquad 
V_z^{ab} = \langle Z_E^{ab} \sigma_z^{ab}\rangle.
\end{equation}
An entanglement witness $W$ can be measured as
\begin{equation}
    W = \sum_{i=1}^{D-1} \sum_{j=i+1}^{D} { N^{ab}  \left( V_x^{ab} + V_y^{ab} + V_z^{ab} \right) },
\end{equation}
where $D$ is the total number of modes (here $D=4$). $N^{ab}$ (corresponding to $|EC\rangle$ state) denotes the normalization constant obtained via projective measurements onto the computational basis, consistent with the $N^{ijk,lmn}$ (corresponding to the $|BCD\rangle$ state) defined in Section~\ref{EW_SM}.

The normalization ensures that each term contributes equally when the state is maximally entangled across the corresponding subspace.

It can be shown that the sum of visibilities over all mode pairs, denoted by $W$, satisfies the following inequality:
\begin{equation}\label{Vio}
W \le \frac{3}{2}D(D-1) - D(D-d),
\end{equation}
where $d$ is an integer. For a given $d$, violation of this inequality implies that the Schmidt number (the genuine entanglement dimensionality) must be at least $d+1$ \cite{Krenn2014,PhysRevA.101.032312}. In other words, the measured $W$ exceeding the right‑hand side of Eq.~\ref{Vio} for a particular $d$ rules out all states with Schmidt rank $\le d$.

The experimentally measured witness value $W = 14.828(99)$ violates the inequality for $d=3$ (the bound for Schmidt number $\le 3$). The uncertainty is estimated from
1000 Monte Carlo samples assuming Poissonian photon-
counting errors. This result unambiguously demonstrating that the bipartite entanglement between $C$ and $BD$ is genuinely four‑dimensional; in other words, the Schmidt rank is no less than $4$. This confirms that the effective four‑dimensional system $E$ shares a full four‑dimensional entangled state with system $C$, despite $E$ being composed of two physical qubits $B$ and $D$.

\end{widetext}
\end{document}